# LIFY: IoT System for Monitoring Vital Signs of Elderly People

S. González[1,2,3], M. Vasquez[1,2], W. Castellanos[1,2]

[1] Ingeniería Electrónica, Pontificia Universidad Javeriana, Bogotá, Colombia.

[2] ARTICO, Semillero de Investigación en Aplicaciones en Red, Telemática e Internet de las Cosas, Pontificia Universidad Javeriana, Bogotá, Colombia.

[3] e – TEAM, Ecosistema - Tecnologías Emergentes para Adultos Mayores, Pontificia Universidad Javeriana, Bogotá, Colombia.

*Abstract* — This article describes the implementation of a technological solution aimed at improving the recording of physiological signals in the elderly population residing in geriatric facilities. The developed system consists of a smart device equipped with sensors for body temperature, heart rate, and blood oxygen levels. This device establishes an Internet connection to transmit data to a cloud-based platform for storage. Within this platform, a dashboard has been created to visualize real-time values captured by the sensors, along with additional functionalities such as user management and the configuration of personalized alerts, which are transmitted to the solution's users through the instant messaging system called Telegram.

## I. Introduction

The importance of the elderly in our society is undeniable. However, they tend to be a forgotten population. As life expectancy increases, this segment of the population becomes more prevalent in society, as shown in the report by DANE (National Administrative Department of Statistics) [1]. This requires the implementation of social policies, healthcare plans for the elderly, as well as technological solutions that enable comprehensive care for this population segment. In many cases, the absence of social programs, alongside factors such as disability and domestic violence, leads to a considerable number of older adults being admitted to geriatric centers. In many instances, they spend their final days facing overcrowded situations, loss of privacy, and impersonal care, as outlined in the study presented in [2].

According to the research presented in [3], in most of the analyzed cases, it was evident that elderly individuals in geriatric facilities are largely tended to by a single person, making it challenging to provide timely and effective care to all patients. This situation leads to various unfavorable conditions. One of these is the inability to maintain an efficient record of the health status of the elderly, as vital signs are taken and recorded manually at different intervals. This hinders the medical monitoring of various conditions and delays the identification of risks for developing new illnesses. Additionally, the limited digitalization of internal processes within geriatric facilities prevents the families of the elderly from accessing health-related information, complicating immediate action in case of an emergency.

Recent advancements in IoT (Internet of Things) have revealed the potential of this technology in various fields, particularly in health monitoring [4]. An example of such applications is described in reference [5]. This solution integrates body temperature, heart rate, and oximetry sensors with embedded systems to implement electronic devices capable of transmitting recorded variables to the internet for remote monitoring. These solutions also offer real-time alerts when a dangerous health condition arises, as shown in the work presented in [6], which sends alert messages directly to a smartphone. Apart from acquiring biomedical signals, other IoT solutions enable the monitoring of environmental variables such as air quality and ambient temperature. This aids in detecting harmful health conditions [7]. These technological solutions allow for more personalized medical care and constant health monitoring, particularly beneficial for patients with chronic and fluctuating needs.

Specifically, IoT platforms focusing on older adults are detailed in references [8, 9], which centrally feature a wearable device that older adults should carry continuously. Other monitoring systems are geared toward equipping homes or geriatric facilities with intelligent components for tracking and monitoring the conditions in which older adults live. Examples of such systems are described in references [10, 11, 12].

In this context, the LIFY system is developed, designed to automate the collection of biomedical signals, digitize the storage process, and integrate a cloud-based platform for continuous monitoring. This system does not only track the medical care staff in the geriatric facility but also provides access to family members. LIFY operates by gathering data, such as body temperature, heart rate, and blood oxygen levels, through an electronic device. These data are transmitted and stored in an internet-hosted application, allowing visualization through a web-based graphical interface. Additionally, automatic alerts are in place to trigger when vital sign alterations occur, sent via text messages using the *Telegram* application. This enables remote monitoring by family members and, in case of significant health indicator changes, the potential to send alerts to medical professionals.

## II. Materials and methods

### A. Materials

In building the biomedical signal acquisition device, two sensors were used: the *GY-906 - MLX90614* infrared temperature sensor and the *MAX30102* heart rate and oxygen concentration sensor. These sensors were interconnected to a Raspberry Pi 3 B+ for data acquisition. For data storage and management, a web application hosted on *Firebase* was developed, and a customized dashboard was created using the IoT platform called *ThingsBoard*. This technological infrastructure allows for the centralized collection and effective presentation of critical health-related information for the elderly. For the final prototype of the device, the parts were designed in 3D and printed in PLA plastic.

### B. Development Phases

The project was structured following the CDIO methodology (Conceive, Design, Implement, and Operate), which comprises four fundamental phases.

In the conception phase, it began with problem research and a review of the current state of the art. The problem was validated through interviews with nurses and personnel responsible for the care of the elderly. Additionally, a visit was made to a facility offering home hospitalization services to identify the needs of the elderly, their families, and the personnel in charge of the geriatric care home or specialized facility.

In the design phase, the selection of sensors, development board, communication protocol, server for the website, dashboard platform, and the type of alarms was conducted.

The implementation phase encompassed the construction of the electronic circuit consisting of the sensors and embedded system, the design and fabrication of 3D components, establishing the internet connection, configuring the cloud platform for data visualization, integrating with the *Telegram* messaging service, setting up alarms, and managing user access.

Finally, in the operation stage, tests were conducted for sensor calibration, verification of automatic alarms, validation of the website's functionality, and access to the dashboards by the respective users.

## III. RESULTS AND DISCUSSION

The architecture of the implemented system can be observed in Fig. 1. The first component is the device responsible for recording biomedical signals. The prototype of this device was built using an embedded system (Raspberry Pi 3 B+) as the processing core. This board is connected to the sensors for temperature, heart rate, and blood oxygen concentration. The physical prototype of the signal acquisition device can be observed in Fig. 2.

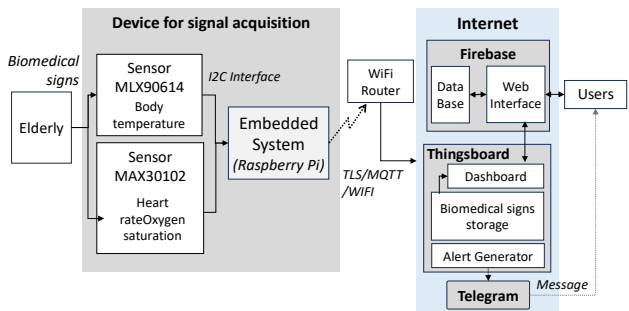

Fig. 1. Block diagram of the main components of the IoT monitoring system – LIFY

The data transmission takes place through a Wi-Fi connection, utilizing the integration of *MQTT* (Message Queuing Telemetry Transport) [13] and *TLS* (Transport Layer Security) communication protocols. *MQTT* is among the most employed protocols in IoT systems due to its low overhead and simplicity, making it suitable for devices with energy consumption constraints. On the other hand, the use of the *TLS* protocol ensures secure data transmission by encrypting the information sent over the internet.

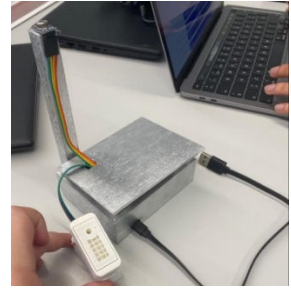

Fig. 2. Prototype of the biomedical signals acquisition device

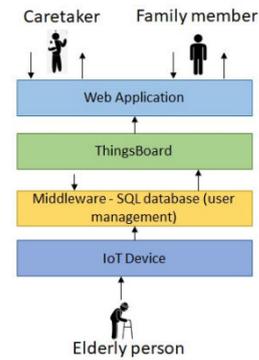

Fig. 3. System Architecture of the Web Application of LIFY

Finally, cloud applications were implemented in Internet in order to provide services as user management, data visualization, alert generator and data storage. The basic architecture of the implemented application can be seen in Fig. 3 and the main services of the web application are summarized in Fig. 4. Firstly, a web application was developed containing general information about the LIFY solution and user management (see Fig. 5). It includes user registration and authentication (login) functions. This way, both the care staff at the geriatric facility and the elderly's family members can access an intranet with personalized information. To ensure individual user management and data privacy, the creation of an account per person on the LIFY platform was established. Upon the initial account creation, users are prompted to fill in details for each elderly individual, considering pre-existing conditions, age, and daily medication, among other relevant data for setting up alarms and monitoring their health.

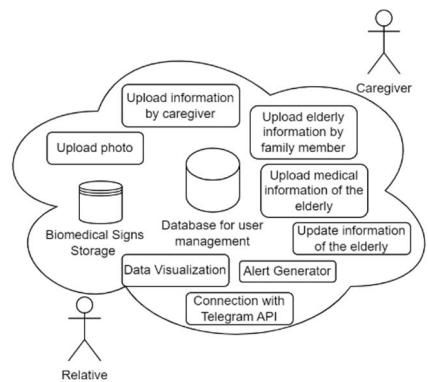

Fig. 4. Services available in the web application

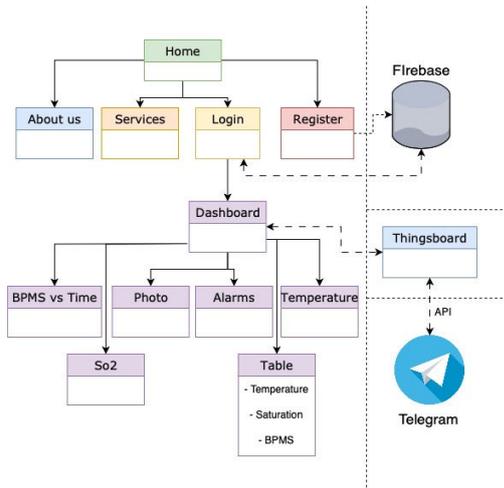

Fig. 5. Structure of the Web Application - https://lify-iot.web.app

The web application was developed using the hosting services, database, and user management features of the *Firebase* platform.

Alongside the web application, a data storage and visualization system was also developed using the *ThingsBoard* platform [13]. Within this platform, the readings of biomedical signals collected by the sensors are stored. Additionally, a custom visualization dashboard was implemented within *ThingsBoard*, allowing the presentation of variables recorded by the sensors through curves and different display types (see Fig. 6). This aids in monitoring health status and identifying abnormal readings by medical staff. Integration with the *Telegram* API was performed on this platform for the automatic sending of alerts upon detecting readings outside the normal range of biomedical signals or when generated manually by the geriatric care staff through the dashboard.

Lastly, integration was established between the web application and the dashboard developed in *ThingsBoard* to facilitate user access to the information within the system.

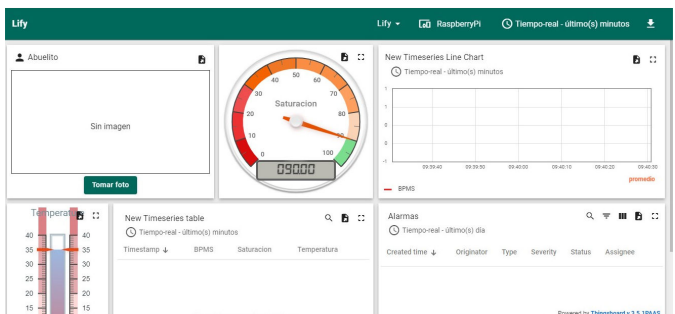

Fig. 6. Dashboard developed in *ThingsBoard*.

IV. CONCLUSIONS

This project stands out for its focus on implementing Internet of Things (IoT) technologies for monitoring the health of the elderly population.

Another remarkable aspect of this project is the personalized automatic alarm system. These alarms are tailored to each patient's specific needs, ensuring timely medical responses in the event of critical health indicator variations. This initiative-taking approach significantly reduces the risk of complications and enhances the quality of life for older adults, providing reassurance to both patients and their caregivers and families.

One of the most significant contributions of the presented work is the ability to access information about the elderly from anywhere using any internet-connected device (smartphone, notebook, or tablet). It is also important to highlight the integration of the cloud platform with the Telegram application, enabling the direct sending of alert messages to the smartphones of the elderly's family members.

What makes this project even more relevant is its focus on cost efficiency. The implementation of low-cost components and technologies, such as the Raspberry Pi, along with free access to Google's *Firebase* platform and the creation of a website on *ThingsBoard*, makes it highly affordable and accessible to a wide range of users. This marks a substantial advancement in democratizing elderly health monitoring compared to similar solutions, which are often expensive and less accessible to the general population.

V. ACKNOWLEDGEMENTS

The authors express their gratitude to Professor Martha Zequera for her support in the project development, the ARTICO research group, and the Faculty of Engineering at Javeriana University for their financial support in the development of this work.